\documentclass[preprint, amssymb]{article}
\usepackage{graphicx}
\usepackage{bm}
\newcommand*{\ZF}{\mathcal Z}
\newcommand*{\VF}{\mathcal V}
\newcommand*{\EF}{\mathcal E}
\newcommand*{\EFB}{\boldsymbol{\mathcal E}}
\newcommand*{\AF}{\mathcal A}
\newcommand*{\AFB}{\boldsymbol{\mathcal A}}
\newcommand*{\MF}{\mathcal B}
\newcommand*{\MFB}{\boldsymbol{\mathcal B}}

\begin{document}

\title{Interferences between En and Mn ($n=1,2...$) optical transition moments}
\author{A. I. Chichinin \\
Institute of Chemical Kinetics and Combustion\\ and
Novosibirsk State University,\\ 
630090 Novosibirsk, Russia\\
chichinin@kinetics.nsc.ru}%

\begin{abstract}
   The operator for interaction between monochromatic electromagnetic field (light)
and molecules is usually presented as a two multipolar series 
expansions $\VF^{E1}+\VF^{E2}+...$ (electro-dipole, electro-quadrupole,...)
and $\VF^{M1}+\VF^{M2}+...$ (magneto-dipole, magneto-quadrupole, ...).
   The optical transition probability is proportional to the complex square of
the sum of these series, which contains interference terms like
$\VF^{En}\VF^{En'}$, $\VF^{Mn}\VF^{Mn'}$, and  $\VF^{Ep}\VF^{Mk}$ ($n\ne n'$).
   In the present study it is shown that all of terms $\VF^{Ep}\VF^{Mk}$
may be observed experimentally, if  the the molecules are oriented,
electrically and magnetically, and the light with appropriate phase shift between
electric and magnetic fields.
   Also, a simple operator for the light-molecule interactions
is proposed, as an alternative to Taylor series mentioned above.
\end{abstract}

\maketitle
\section{Introduction}
   The theory of interactions between light and molecules (or atoms)
usually employs interactions of the electric and the magnetic
fields of the light with a multipolar moments of the molecule\cite{Barron04,Raab05}.
   Thus, electric field of light $\EFB$ produces electro-dipole (E1),
electro-quadrupole (E2), electro-octupole (E3) \emph{etc.} interactions.
   We will call them different types of interactions.
   Mathematically, these interactions comes from the Taylor series expansion:
\begin{equation}
\label{I0}
e^{i \theta } \approx 1 +i\theta  - \frac{1}{2} \theta^2 +\dots,
\end{equation}
where $\theta \equiv  {\bm k} \cdot {\bm r} \ll 1$ is assumed.
   Hereafter ${\bm k}$  is a wave vector, pointing in the direction of light propagation,
and ${\bm r}$ is a vector from center of molecule to the electron.

  The interactions of magnetic field $\MFB$ with magnetic multipoles of the molecule.
are described similarly.
  The result is the consequence of
magneto-dipole (M1), magneto-quadrupole (M2), magneto-octupole (M3) \emph{etc.} interactions.

   In the following, we consider absorption linearly-polarized-light spectroscopy
of molecule (or atom) with one electron.
   The extension to emission spectroscopy, to different polarizations of light,
and to many-electron systems is straightforward.

   The rate of the optical transition from state $i$ to state $f$ in the molecule
is usually calculated in the first order of perturbation theory as a square of
transition matrix elements (Fermi's "Golden rule"):
\begin{equation}
\label{II}
W_{i\to f} = \underbrace{\frac{2\pi}{\hbar^2} \delta(\omega- \omega_{i\to f})}_{\equiv C}\;%
  | \sum_{n=1}^{\infty} \left( \hat{\VF}^{(E_n)}_{fi} +\hat{\VF}^{(M_n)}_{fi} \right) |^2,
\end{equation}
where $\omega$ and $\omega_{i\to f}$ are probe-light and molecule-transition
frequencies, respectively;
the time-independent operators of the electrical and magnetic interactions
are denoted as $\hat{\VF}^{(En)}$ and $\hat{\VF}^{(Mn)}$, respectively.
  Hereafter subscript $fi$ denotes transition matrix element,
$A_{fi} \equiv \langle f|A|i \rangle$,
and subscript $i\to f$ is index only.
  The prefactor $C$ hereafter may be treated as a constant.
   All of the other characteristic of the transition like:
oscillator strength $f_{i\to f}$, absorption cross section $\sigma_{i\to f}$,
line strength $S_{i\to f}$, and rate of spontaneous decay $A_{i\to f}$ are proportional
to the speed $W_{i\to f}$.

  Rearrangement of Eq. (\ref{II}) gives
\begin{equation}
\label{I1}
\frac{W_{i\to f}}{C} =  \sum_{n=1}^{\infty} \left(  |\hat{\VF}^{(E_n)}_{fi}|^2+ |\hat{\VF}^{(M_n)}_{fi}|^2 \right) + I,
\end{equation}
where $I$ is the sum of all interference terms (ITs):
\begin{eqnarray}
\label{I3}
\nonumber
I  &\equiv& \sum_{n \ne n',m \ne m'}  \hat{\VF}^{(E_{n})*}_{fi}\hat{\VF}^{(E_{n'})}_{fi} +
\hat{\VF}^{(M_{m})*}_{fi}\hat{\VF}^{(M_{m'})}_{fi}+ \\
&+& \hat{\VF}^{(E_{n})*}_{fi}\hat{\VF}^{(M_{m'})}_{fi}+
\hat{\VF}^{(M_{m})*}_{fi}\hat{\VF}^{(E_{n'})}_{fi}.
\end{eqnarray}
   Usually, only the largest term in sum (\ref{I1}) is used
and all the ITs from Eq. (\ref{I3}) are neglected.
   One term in sum (\ref{I1}) usually is enough because of
strong hierarchy between probabilities
of transitions of different types:
\begin{eqnarray}
\nonumber
&&E1:E2:E3:M1:M2:M3 =\\
\nonumber
&&= 1:(k\rho)^2:(k\rho)^4:
(\ZF\alpha)^2:(\ZF \alpha)^2(k\rho)^2:(\ZF\alpha)^2(k\rho)^4,
\end{eqnarray}
where $\rho$ is a typical size of molecule,
$\alpha= e\hbar/m_ec \approx 1/137$ is a fine-structure constant,
$k=|{\bm k}|$, and $\ZF$ is nuclear charge. 

   Recently, the interest to sum (\ref{I1})  was recommenced.
   One reason is the X-ray spectroscopy, where the characteristic wavelength
is $\sim$ 1--10~\AA, and hence the long-wave approximation (\ref{I0})  is no longer
valid\cite{George07_965,Bernadotte12_204106,Lestrange15_234103}.

   Another reason is the "origin problem": truncating of series (\ref{I0})
may result in a pronounced origin dependence of transition probability (\ref{I1}).
   In pother words, the calculated E$n$- and M$n$-transition probabilities may
strongly depend on the position of the coordinates
origin\cite{George07_965,Bernadotte12_204106,Lestrange15_234103}.
   Even negative oscillator strengths may be obtained\cite{Lestrange15_234103}.

   The purpose of this work is twofold: first, to
propose an experiment, in which the interference between different types
of interactions may be observed.
   Such experiment could be important prove of basics of quantum mechanics.
   To the knowledge of the author there has been no such experiment.

   And second purpose is to propose analytical expression for light-matter interactions
to be used instead of series (\ref{I1}).
   The expression may be used by computing programs (Molpro, Gaussian, Dalton,...),
which predict characteristic of the optical transitions.
   Also, the expression may be very useful in the computational
studies, where contributions of different transition types are
analyzed and compared\cite{Safronova06_47}.

   In the literature there are different types of spectral interferences.
   One kind is, for example, a mutual interference between
spectral lines of atomic Ga and Mn, because their wavelengthes
are very close (4032.98 and 4033.07 \AA, respectively)\cite{Allan69_13}.
   Another kind of interferences originate from
mixing of two close states of molecule.
   For example, in spectroscopy of diatomic molecules
it is called "quantum mechanical interference effect" \cite{Lefebvre-Brion04}.
   Another well known name of this effect is "Fermi resonance".
   Important, that in the present study we deal with a new type of spectral interference.

\section{Obstacles preventing observation of ITs}
\subsection{Multipole operators}
   A plane light wave may be described by well-known equations:
\begin{eqnarray}
\label{I5c}
\AFB({\bm r},t) &=& i \AF_0 {\bm e}_{\EF} e^{i({\bm k} \cdot {\bm r} - \omega t)},\\
\nonumber \tilde{\EFB}({\bm r},t) &=& -\nabla \phi - \frac{1}{c}\; \frac{\partial \AFB}{\partial t} =
k \AF_0 {\bm e}_{\EF} e^{i({\bm k} \cdot {\bm r} - \omega t)}=\\
\label{I5a}
   &=& \EFB({\bm r})e^{-i \omega t},\\
\nonumber
\tilde{\MFB}({\bm r},t) &=& \nabla \times \AFB =
- \AF_0 ({\bm k}\times {\bm e}_{\EF}) e^{i({\bm k} \cdot {\bm r} - \omega t)}=\\
\label{I5b}
   &=& \MFB({\bm r})e^{-i \omega t},
\end{eqnarray}
where $\AFB({\bm r},t)$ is vector-potential and
$\phi$ is the scalar potential (which for light normally is put to zero).
   Vectors $\EFB$ and $\MFB$ are time-independent electric and magnetic fields of light,
and ${\bm e}_{\EF}= \EFB/|\EFB|$ and ${\bm e}_{\MF}= \MFB/|\MFB|$ are their unit vectors,
respectively.
   The operator for non-relativistic interaction of light with the molecule is given by expression:
\begin{equation}
\label{LA1_general}
\hat{U} = -\frac{e}{m_e c} (  \hat{\bm p} \cdot \AFB({\bm r},t) ) - \mu_B %
 (\MFB({\bm r},t) \cdot \hat{\bm S}),
\end{equation}
where $\hat{\bm p}$ is the momentum operator of the electron,
$\hat{\bm S}$ is dimensionless  operator of spin,
$\mu_B \equiv e\hbar/2m_ec$ is Bohr magneton,
and $g_e \approx 2.00232$ is $g$-factor of free electron.
   The factor $e^{-i \omega t}$ in expressions (\ref{I5c}--\ref{I5b}) is used to obtain
Fermi's "Golden rule" (\ref{II}),
and the factor $e^{i {\bm k} \cdot {\bm r}}$ produces Taylor series (\ref{I0}).

   The operators for the multipole interactions are presented in Table \ref{tab:T1}.
   In the literature there are several definitions of magnetic moments,
we prefer the approach of Raab\cite{Raab75_1323}.

   We just simplified formulae of Raab,
writing $ 2 \theta^k \hat{\bm L}$ instead of original hermetian
$ \theta^k \hat{\bm L} + \hat{\bm L} \theta^k $,
because we assume communication rule
${\bm e_{\MF}} \cdot \hat{\bm L} \theta = {\bm e_{\MF}}  \cdot \theta \hat{\bm L}$.
   This rule follows from the relation
$ \hat{L}_y \theta  - \theta  \hat{L}_y + i k{r}_x$;
hereafter right-handed cartesian axis system
$({\bm e}_{x},{\bm e}_{y},{\bm e}_{z})=({\bm e}_{\EF},{\bm e}_{\MF},{\bm e}_{\bm k})$ is used.

   The $\hat{\VF}^{(En)}$ operators are presented here only in distance form,
they have opposite sign in comparison with formulae of Bernadotte
\emph{et al.}\cite{Bernadotte12_204106},
because we calculate the matrix elements $\hat{\VF}_{fi}$, and they calculated
the matrix elements $\hat{\VF}_{if}$.

\begin{center}
\begin{table}[h]
\caption{\label{tab:T1}%
  Operators of light-molecule interactions,
first terms of E$n$ and M$n$ series\cite{Bernadotte12_204106,Raab75_1323},
and general expressions.
   Magnetic and electric fields here are synchronous,
$\MFB({\bm r},t) = ( {\bm k} / |k| ) \times \EFB({\bm r},t)$.
   Expressions in square brackets are electric and magnetic moments of the molecule.}
\begin{tabular}{l  ll}
\hline
$\hat{\VF}^{(E1)}=$& $\EF_0\;  {\bm e}_{\EF} \cdot [e {\bm r}]_{fi}\; ^a$
 & \\
$\hat{\VF}^{(E2)}=$& $\EF_0\; (i/2!)\; ({\bm e}_{\EF} \cdot  [e {\bm r} \theta]_{fi}$ \\
$\hat{\VF}^{(E3)}=$& $\EF_0\;  (i^2/3!)\; {\bm e}_{\EF} \cdot [e {\bm r}\theta^2]_{fi}$ \\
$\hat{\VF}^{(En)}=$& $\EF_0\; (i^{n-1}/n!)\; {\bm e}_{\EF} \cdot [e {\bm r}\theta^{n-1}]_{fi}$\\
\hline
$\hat{\VF}^{(M1)}= $& $\MF_0\; {\bm e}_{\MF} \cdot [ \mu_B (\hat{\bm L} + g_e \hat{\bm S})]_{fi}\; ^b $
& \\
$\hat{\VF}^{(M2)}=$ & $ \MF_0\; (i/2!)\; {\bm e}_{\MF}\cdot %
[2\theta  \mu_B  (\frac{2}{3} \hat{\bm L} + g_e \hat{\bm S})]_{fi} \; ^c$   \\
$\hat{\VF}^{(M3)}=$ & $ \MF_0\;  (i^2/3!)\; {\bm e}_{\MF}\cdot[ %
3 \theta^2  \mu_B ( \frac{ 2}{4} {\bm \hat{\bm L}} + g_e \hat{\bm S} ) ]_{fi} $ \\
$\hat{\VF}^{(Mn)}=$ & $ \MF_0\; (i^{n-1}\!/n!)\; {\bm e}_{\MF} \cdot %
[ n \theta^{n-1}  \mu_B (\frac{2}{n+1} \hat{\bm L} + g_e \hat{\bm S} ) ]_{fi} $ \\
\hline
\end{tabular}\\
$^a$ $ \theta \equiv ({\bm k} \cdot {\bm r})$,
$ \EF_0 = \MF_0 \equiv k \AF_0$,
${\bm e}_{\EF}= \EFB/|\EFB|$,
and $k = |{\bm k}|=\omega/c$.\\
$^b$ ${\bm e}_{\MF}= \MFB/|\MFB|$,
$ \hat{\bm L} \equiv ({\bm r} \times {\bm p})/\hbar$, and ${\bm S}$ is dimensionless.\\
$^c$ Misprint in Ref. \cite{Bernadotte12_204106} here: factor 2 before $\hat{\bm S}$ is lost.
\end{table}
\end{center}

\subsection{ITs vanish due to $i$-shift}
   Now we start specify the obstacles, which prevent observation of ITs.

   The most evident reason for zero IT between interactions $O$ and $O'$ is
the phase shift between them of $\pm i = e^{\pm i\pi/2}$:
\begin{equation}
\label{I4}
O^*_{fi} O'_{fi} + O_{fi} O'^*_{fi} = 2 Re(O^*_{fi} O'_{fi})=0.
\end{equation}
   Hereafter we call it $i$-shift.
   For example, it is clear from Eq. (\ref{I0}), that
interference between E1- and E2- interactions is impossible. 

   More generally, interference between E$n$ and E$n'$ terms
(as well as between M$n$ and M$n'$ terms) is possible only if
both numbers $n$ and $n'$ are even or both of them are odd.
   The interference between E$n$ and E$n'$ terms,
where $n$ and $n'$ have different parities, is impossible,
because there is the $i$-shift between them.

\subsection{Vectors ${\bm r}$ and ${\bm L}$ anisotropy}
   Some  ITs, like E1-E2, are usually neglected,
because their operator is proportional to $({\bm k} \cdot {\bm r})^n$ where $n$ is odd,
because they vanish after averaging over directions of vector ${\bm r}$.
   But it is easy to get read of this limitation.
   One can orient molecules by means of
static electric fields\cite{Loesch90_4779,Holmegaard10_428} or optical
fields\cite{Stapelfeldt03_543,Thomann08_9382,Goban08_013001,De09_153002},
or both of them\cite{Friedrich99_10280,Ghafur09_289,Hansen13_234313}.
   For example, in experiments of Hansen {\emph{et al.}} the laser-induced
one-dimensional orientation $ \langle \cos \theta_z  \rangle \approx 0.99$ of
CPC (C$_4$N$_2$H$_2$ClCN) molecules have been demonstrated,
as well as the three-dimensional orientation of these molecules\cite{Hansen13_234313}.
   In other words, if we want to observe the ITs experimentally,
the anisotropy of vector ${\bm r}$ is an overcomable obstacle.

   If we need to orient magnetic moment in molecule, it also
may be done by strong magnetic field.

   The oscillating character of electric and magnetic fields
is also not a problem, because in normal light they oscillate synchronously,
see Eqs. (\ref{I5a},\ref{I5b}).
   For example, in order to obtain observable IT between E1 and M1 transitions,
we orient dipole and magnetic moments of the molecules
by two strong constant fields, electric and magnetic, respectively.
   From classical point of view, the
electro-dipole interaction $({\bm r} \cdot \tilde{\EFB}({\bm r},t))$
oscillate as $\cos \omega t$,
and magneto-dipole interaction $({\bm r} \cdot \tilde{\MFB}({\bm r},t))$
oscillate as $\cos \omega t$,
but their product oscillate as $\cos^2 \omega t$.
   Therefore, the time-averaged value of the product is not zero.

\subsection{Quantum numbers for E-M ITs  }
   \emph{1. Selection rules for ITs between En and Mn'. }

   According to Wigner-Eckart theory, if light electric field is directed along the
axis of quantization, $\EFB \parallel {\bm e}_z$,
the selection rule for each electro-multipole interaction is $\Delta M_J=0$:
\begin{equation}
\langle J_f,M_f | r^nC^{n}_0 |J_i,(M_i)  \rangle
= (-1)^{J_f-M_f}
\left(\begin{array} {ccc}J_f&n&J_i\\ M_f&0&M_i \end{array}\right) %
\langle J_f|| r^n C^{n} ||J_i \rangle,
\end{equation}
where $J_i,J_f$ and $M_i,M_f$ are respectively molecular total rotational moments
and their projections.
   Here electric multipole moment has form
$- r^n C^{n}_0({\vartheta,\phi})$,  where $C^{n}_m({\vartheta,\phi})$
is spherical unit vector of Weissbluth.

   If we want to observe ITs between electric and magnetic matrix elements,
the states $f$ and $i$ should be the same for both matrix elements.
   Hence the selection rule for magnetic matrix element should be also $\Delta M_J=0$,
leading to the condition $\EFB \parallel \MFB \parallel {\bm z}$,
which for light is impossible.

   Therefore, we should consider only $\Delta M_J=\pm 1$ selection rule for perpendicular
transitions, $\EFB \perp {\bm z}$ and  $\MFB \perp {\bm z}$.

   \emph{2. Difference between En and M1 transitions. }

   Let us imagine NO molecule which is oriented, electrically and magnetically,
 along axis $z$, which coincides with light vector ${\bm k}$.
   Magnetic field of the light can produce M1-transition,
due to matrix element $\langle   ^2\Pi_{1/2} |\hat{S}_y | ^2\Pi_{3/2} \rangle$,
see, for example, detection of BrO radical\cite{McKellar81_43}.

   As it was mentioned above, the light electric field should produce the same perpendicular
transition, but the molecule has no dipole moment which is
perpendicular to the axis of molecule.
   Hence matrix element $ \langle ^2\Pi_{1/2} |{\bm r}_x | ^2\Pi_{3/2}  \rangle$ is zero.

   In general, M1-magnetic transitions normally just change
mutual direction of $L$ and $S$, and therefore only the total moment ${\bm J}$
ant it's projections ($M_J$, $\Omega$) can be changed.
   Conversely, E1-magnetic transitions normally change
some other quantum numbers, in addition to the numbers $J$, $M_J$, and $\Omega$.

   This incompatibility looks very serious for all molecules, hence probably
only atoms can show non-zero ITs between E$n$- and M1-excitations.

\subsection{Phase shift between electric and magnetic fields}
   Let us consider an experimental detection of the E1-M1 interference term
$({\bm r} \cdot \EFB)_{fi}^*\; ({\bm \mu} \cdot \MFB)_{fi}$.
   As it is shown above, an atom should be oriented, electrically and magnetically,
 along axis $z$, and  ${\bm z} \parallel {\bm k}$.
   The IT may be rearranged to the form $({r}_x)_{fi}^*\; (\mu_y)_{fi}$.
   There is no $i$-shift between these two matrix elements.

   However, interference is impossible.
   This restriction comes from the fact, that the matrix elements of $x$-oriented vectors
differ from the matrix elements of $y$-oriented vectors by imaginary unit $i$.

   This may be shown from the transformation matrix\cite{Varshalovich88}:
\begin{equation} \label{LA2_unit_vektor}
\left( \begin{array}{c} e_x \\ e_y \\ e_z \end{array} \right) =
\left( \begin{array}{ccc}
-1/\sqrt{2} &1/\sqrt{2} &0 \\
 i/\sqrt{2} & i/\sqrt{2} &0  \\
0 &0 &1 \\
\end{array} \right)
\left( \begin{array}{c} e_{1} \\ e_{-1} \\ e_{0} \end{array} \right),
\end{equation}
which may be obtained from the properties of the spherical $Y_{1m}$-functions.
  Here  $e_x$, $e_y$, and $e_z$ are cartesian components of a unit vector,
and $e_1$ $e_0$ $e_{-1}$ are spherical coordinates of the vector;
in our case $e_x = e_{\EF}$ and $e_y=e_{\MF}$.

   It means, that the term
$({\bm d} \cdot \EFB)_{fi}^*\; ({\bm \mu} \cdot \MFB)_{fi}$
will be imaginary in comparison with $|{\bm d} \cdot \EFB|^2_{fi}$ and
$|{\bm \mu} \cdot \MFB|^2_{fi}$.
   Therefore, for synchronous magnetic and electric fields of light,
IT for E1-M1 interactions is impossible because of the right angle between the
electric and magnetic fields.

\subsection{Phase shift between M1 and E2 interactions}
   There is a $i$-shift between M1 and E2 operators,
see Table~\ref{tab:T1}.
   It may be shown, for example, by a very straightforward proof
in the work  of Bernadotte \emph{et al.}\cite{Bernadotte12_204106},
where both M1- and E2- operators are derived from one origin.

   However, according to arguments from the previous section,
there is the second $i$-shift due to right angle between vectors $\MFB$ and $\EFB$.
   There are two shifts, and therefore, the IT between M1- and E2- interactions is possible.

\subsection{Summary of the obstacles}
   The obstacles preventing IT observation are listed in Table~\ref{tab:T2}
for interactions up to $n=3$.

\begin{center}
\begin{table}[h]
\caption{\label{tab:T2}%
The obstacles preventing IT observation between different interactions,
for normal light with synchronous electric and magnetic fields.
  If the phase between electric and magnetic fields of light is set to $\pi/2$,
all the E-M ITs should be multiplied by $i$.}
\begin{tabular}{cc cc cc c}
\hline
   & E1 & M1 & E2 & M2 & E3 & M3 \\
\hline
E1 & A $^a$  & %
i$^b$ $\!{\bm R_l}^c {\bm S}^d$  &$i ^e {\bm R_k}\!^f$ &  ${\bm R_l}{\bm R_k}{\bm S}$ & A  & i${\bm R_l}{\bm S}$  \\
M1 &    & A  & ${\bm R_l}{\bm R_k}{\bm S}$ &
$i{\bm R_k}$%
                  &i${\bm R_l}{\bm S}$  &  A  \\
E2 &    &    & A  & i${\bm R_l}{\bm S}$ & $i{\bm R}_k$ & ${\bm R_l}{\bm R_k}{\bm S}$ \\
M2 &    &    &    &  A & ${\bm R_l}{\bm R_k}{\bm S}$  & $i{\bm R_k}$  \\
E3 &    &    &    &    & A  & i${\bm R_l}{\bm S}$  \\
M3 &    &    &    &    &    & A  \\
\hline
\end{tabular} \\
\renewcommand{\baselinestretch}{0.75}
\small
 $^a$: {A $\equiv $ allowed}.\\
 $^b$: {i means 90$^{\circ}$ phase  shift due to $\EFB \perp \MFB$.}\\
 $^c$: {${\bm R_l} \equiv $ isotropy of ${\bm r}$ along vector $\EFB$.}\\
 $^d$: {${\bm S} \equiv $ isotropy of vectors ${\bm L}$ and ${\bm S}$ along vector $\MFB$.}\\
 $^e$: {$i$ meas phase shift between O$n$ and O$n+1$ operators}\\
 $^f$: {${\bm R_k} \equiv $ isotropy of ${\bm r}$ along vector ${\bm k}$.}
\normalsize
\renewcommand{\baselinestretch}{1}
\end{table}
\end{center}
   In summary, there are four groups of terms:
E$^en$, E$^on$, M$^en$, and M$^on$; hereafter indexes $o$ and $e$ denote parities
of $n$ numbers.

   There ITs inside each group exist,
but ITs between the members of different groups is equal to zero
except the E$^on$-M$^en'$ and E$^en$-M$^on'$ ITs which in principle may be detected
in oriented atoms.

\section{How to detect E-M ITs }
\subsection{M1--E2}
\begin{figure*}
 \includegraphics[width=\textwidth]{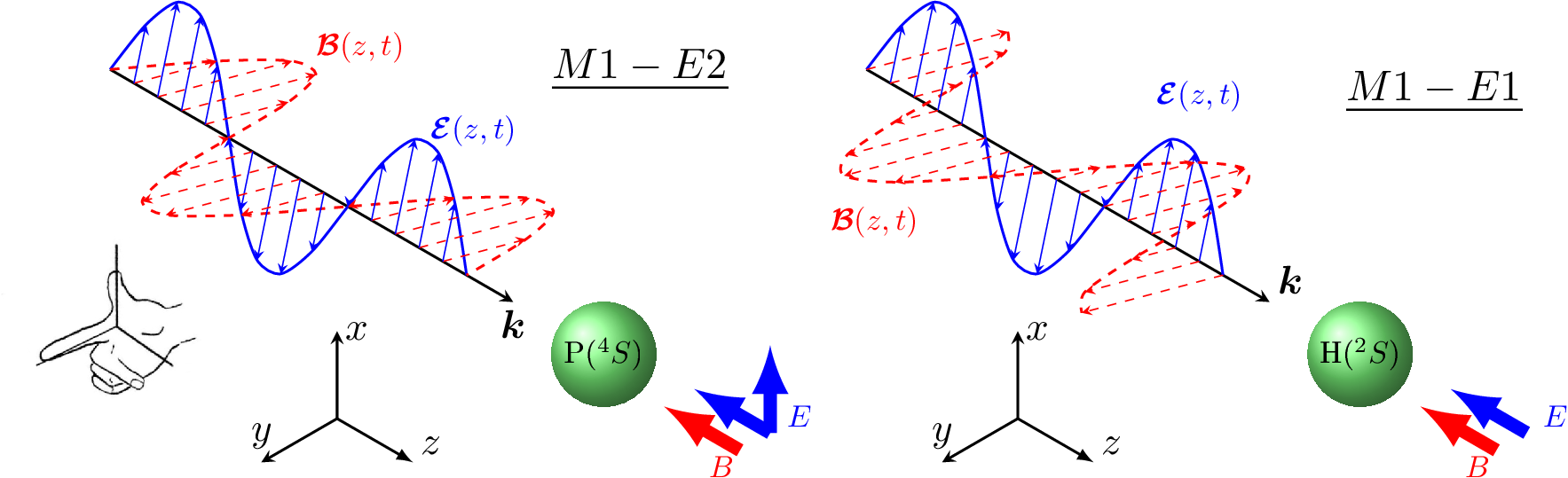}
 \caption{\label{fig:Interfer} Principal scheme of experiments
 in which  M1--E2 and M1--E1 ITs could be observed.
 Left: M1--E2 experiment with phosphorus atoms.
 No phase shift between electric ($\EFB(z,t)$) and magnetic ($\MFB(z,t)$)
 fields of radiation, static electric field $E$ has two components.
 Right: M1--E1 experiment with hydrogen atoms.
 There is a phase shift between electric and magnetic fields.}
 \end{figure*}

  The most promising candidate for IT-detection experiment
may be M1- and E2- transitions.
   In order to observe them, an atom with close
probabilities of E2- and M1-transitions should be chosen.
   The dipole moment of the atom should be oriented
along two axis, $\EFB$ and ${\bm k}$,
and the magnetic moment should be oriented along the axis $\MFB$,
see fig. \ref{fig:Interfer}.

   It may be, for example, the phosphorous atom
at the transition $3s^2 3p^3[X ^4\!S^{\circ}_{3/2} -^2\!\!D^{\circ}_{\circ} ]$
near 880 nm, which has emission inverse lifetimes of
$1.77\times10^{-4}$ and $1.2\times10^{-4}$  s$^{-1}$ for M1- and E2-components,
respectively\cite{NIST_ASD}.
   Note that the $X ^4S^{\circ}_{3/2}$ state is the ground state of the atom,
this fact facilitates experimental preparation of the atom.

\subsection{Phase shift between $\EFB$ and $\MFB$}
   Table~\ref{tab:T2} is not the end of a story.
   Equations (\ref{I5c}--\ref{I5b}) tell us, that the phase shift $\Delta \phi$ between
electric wave $\EFB$ and magnetic wave $\MFB$ of light
far from the source of the light (antenna)  is zero, $\Delta \phi=0$.

   However, the phase shift $\Delta \phi$
 vary with distance between antenna and observation point.
   Close to the antenna, $|\Delta \phi| \approx \pi/2$.
   As the distance from the antenna increases, the phase $\Delta \phi $ decreases.
   This is the basic principle which is used today in near-field electromagnetic
ranging (NFER) technology.

   Moreover, in electron paramagnetic resonance (EPR)
the microwave resonator is used, where electric and magnetic field
components are exactly out of phase, $|\Delta \phi| = \pi/2$.

   Therefore, if one uses such out-of-phase radiation,
all the E-M interactions in Table \ref{tab:T2}  should be multiplied by
imaginary unit $i$.

\subsection{E1-M1}
   If EPR microwave resonator is available,
observation of  IT between M1- and E1-transitions becomes possible.
  In order to observe them, one should orient both the magnetic moment
and the dipole moment  of the atom  along axis $z$, see fig. \ref{fig:Interfer}.

  The simplest system here is probably hydrogen atom H($^2\!S$).
  With these two orientations, very simplified wave functions of
the initial and the final states of the atom becomes
\begin{eqnarray}
\nonumber
\Psi_i &=& a \Psi(^2\!S,-1/2)+b \Psi(^2\!P,-1/2),\\
\nonumber
\Psi_f &=& a'\Psi(^2\!S,1/2)+b' \Psi(^2\!P,1/2),
\end{eqnarray}
where we denoted $(^2\!S,\pm 1/2) \equiv (1s[^2\!S], m_z=\pm 1/2)$,
$(^2\!P,\pm 1/2) \equiv$ $(2p[^2\!P_{1/2,3/2}]$, $m_z=\pm 1/2)$,
and $a$, $b$, $a'$,  and $b'$ are real coefficients.
   In reality, orientation of atom by constant electric field produces
a sum of different antisymmetric states of the same parity, not only  $2p[^2P]$.

   In order to increase population difference between two Zeeman-split
$m_z=\pm 1/2$ sublevels, low temperature should be employed.
   Here for simplicity we assume so low temperatures,
that contribution of $|1s[^2S], m_z=+1/2>$ state to $\Psi_i$ is negligible.

   The IT may be calculated as:
\begin{eqnarray}
\nonumber
\!\! ({\bm d} \cdot \EFB)_{fi} ({\bm \mu} \cdot \MFB)_{fi}\! &\!=\!&
\! \EF_x\! \left[ ab'   \langle ^2\!P,\frac{1}{2}| d_x|^2\!S,-\frac{1}{2} \rangle +
ba'  \langle ^2\!S,\frac{1}{2}|d_x|^2\!P,-\frac{1}{2} \rangle \right] \times \\
&\! \times \!&\! \MF_y \! \left[aa'  \langle ^2\!S,\frac{1}{2}|i \mu_y|^2\!S,-\frac{1}{2} \rangle +
bb'  \langle ^2\!P,\frac{1}{2}|i \mu_y|^2\!P,-\frac{1}{2} \rangle \right]\!,
\end{eqnarray}
where $\EF_x$ and $\MF_y$ are real values.

  Important, that this expression is real,
this fact makes it possible to contribute to
E1- and  M1- transitions, which are proportional to
$\EF^2_x|{\bm d}_{fi}|^2$ and $\MF_y^2 |{\bm \mu}_{fi}|^2$, respectively.

\section{Analytical summation of multipole series}
   Summarizing operators for multipole interactions from Table \ref{tab:T1},
we receive a new method for calculation of light-molecule interactions,
which does not require the summation of endless series (\ref{I1}):
\begin{equation}
\label{Main_Sum}
\frac{W_{i\to f}}{C} =|V^{Eo}_{fi}|^2+|V^{Ee}_{fi}|^2+ |V^{Mo}_{fi}|^2+|V^{Me}_{fi}|^2,
\end{equation}
where
\begin{eqnarray}
  V^{Eo}&\equiv& \sum_{n=0}^{\infty} \hat{\VF}^{(E2n+1)}= %
  e \frac{ \sin\theta}{\theta} \; ({\bm r} \cdot \EFB),\\
V^{Ee}&\equiv& \sum_{n=1}^{\infty} \hat{\VF}^{(E2n)}  = %
  ie \frac{(1-\cos\theta)}{\theta} \; ({\bm r} \cdot \EFB), \\
\nonumber V^{Mo}&\equiv& \sum_{n=0}^{\infty} \hat{\VF}^{(M2n+1)}  = %
 \mu_B \left[ \frac{2(\cos \theta +\theta \sin\theta-1)}{\theta^2} \hat{\bm L}+ \right. \\
 \label{Mo}
   &+& \left. \cos\theta\; g_e \hat{\bm S} \right] \cdot \MFB,\\
\nonumber V^{Me}&\equiv& \sum_{n=1}^{\infty} \hat{\VF}^{(M2n)}= %
 i \mu_B \left[ \frac{2 (\sin\theta - \theta \cos\theta)}{\theta^2} \hat{\bm L}+ \right. \\%
 \label{Me}
    &+& \left. \sin\theta\; g_e \hat{\bm S} \right] \cdot \MFB.
\end{eqnarray}

   If necessary, both $V^{M}$ expressions may be easily transformed to hermetian form
via replacing of $f(\theta)\hat{{\bm L}}$ by $[f(\theta)\hat{{\bm L}}+\hat{{\bm L}}f(\theta)]/2$,
where $f(\theta)$ denote expressions before ${\bm L}$ in Eqs. (\ref{Mo}) and  (\ref{Me}).

  It is well known that calculations of truncated Taylor expansion (\ref{I0})
may lead to wrong results.
   According to George {\emph{et al.}}, "...the calculated E2- and M1-transition
probabilities may grow to arbitrarily large values depending on the position of the
coordinates origin"\cite{George07_965}.

   If one uses our expressions (\ref{Main_Sum}), it is evident,
that the "origin problem" probably remains, but large values of transition
probabilities cannot appear, because all the functions
of variable $\theta$ quickly decrease with $\theta$.

\section{Results and conclusions}
   In summary, we classified the
obstacles which prevent observation of interference terms.
  These terms arise in the calculation of light-matter interactions,
if these  interactions are presented as a series of
interactions between multipole moments of the system (atom or molecule)
and electric or magnetic fields of the probe electromagnetic radiation,
see Table \ref{tab:T1}.

   Some of these obstacles are shown to be overcomable,
 see Table \ref{tab:T2}, and we pointed out the ways, how to do it.
    Thus, orientation of atoms (or molecules) should be used and,
 in some cases --- appropriate phase shift between electric and magnetic fields
 of the probe radiation.
    As a result, we proposed experimental schemes which could
 be used to observe E1-M1 and E2-M1 interference terms.

   Also, we propose simple analytical expression (\ref{Main_Sum}) for light-matter interactions
to be used instead of the series of multipole interactions
and interferences between them.
   Hopefully, this expression will find application in the X-ray spectroscopy.

\bibliographystyle{unsrt}

\end{document}